\documentclass{article}
\usepackage{amsmath}
\usepackage{amsfonts}
\usepackage[english]{babel}
\usepackage{graphicx}
\usepackage[latin1]{inputenc}
\title{A structural analysis on Gravity of Trade: on removing distance from the model}
\author{Rodolfo Metulini}
\begin{document}
\maketitle

\begin{abstract}
The Gravity Model is the workhorse for empirical studies in International Economies for its empirical power and it is commonly used in explaining the trade flow between countries; it relies on a function that relates the trade with the masses of the two countries and the distance (as a proxy of the trasport costs) between them. However, the notion that using of distance functions in conventional interaction models effectively captures spatial dependence in international flows has long been challenged. It has been recently fully recognized that a spatial interaction effect exists essentially due to the spatial spillover and the third country effect. This motivates the introduction of the spatial autoregressive components in the so-called spatial gravity model of trade. A so-called weight matrix is used in order to define the set of the spatial neighbors and it is traditionally based on the inverse of the distance. Two issues follow from this standard procedure: the first regards the biasness of the distance if it is used as a proxy of the transport costs in a panel data, the second is related to the collinearity emerging if we use distance twice. So, several attempt were made in the recent literature having the scope of remove the distance. We propose a theoretically consistent procedure based on Anderson, Van Wincoop derivation model, and some ad-hoc tests, relating to this attempt. The empirical results based on a 22-years panel of OECD countries are conforting, and they allow us to estimate the model without the distance, if properly replaced by a set of fixed effects. This article, in addition, fits in the dispute about how to estimate the multilateral resistance terms.
\end{abstract}
\section{Introduction}

The literature shows how, at empirical level, the classical Gravity Model (Anderson 1979, Anderson, Van Wincoop 2003) brings to  good results in explaining international trade.   Anderson (1979, p.106) states that this equation is the most successful in explaining this issue, furthermore, Everett and Hutchinson (2002, p-489) define this model as the workhorse for empirical studies in international economy. Since the gravity model have physical roots, the trade flow depends on the economic dimension of both the origin and the destination countries, as well as the trade costs between them.

From the beginning of the gravity model of trade, trade costs was taken into account using the distance variable and other variables considered relevant on the volume of trade, such as geographic contiguity, common language and common currency, and the presence of some Free Trade Agreements.

However, some discoveries on the gravity model of trade in the recent literature are relevant, and can be summarized as follow:
\begin{itemize}
\item The original model by Anderson did not considers some effects related to the (auto)correlation in space: the notion that using of distance functions in conventional interaction models effectively captures spatial dependence in international flows has long been challenged. Starting from the Tobler first law of Geography (1970) \footnote{everything is related to everything else, but closer things are more closely related than distant things}, it has been recently fully recognized that a spatial interaction effect exists (Bang, 2006; Kelejian, Tavlas, Petroulas, 2012; Baltagi, Egger, Pfaffermayr, 2007; Hall, Petroulas, 2008), essentially due to the spatial spillover and the third country effect. This motivates the introduction of the spatial autoregressive components in the so-called spatial gravity model of trade. The spatial gravity model of trade make use of the spatial econometric techiques (Anselin, 1988), developed for flow data by LeSage, Pace (2008) and Fischer, Griffith (2008). 

\item Moreover, Anderson and van Wincoop (2003) derive a reduced-form gravity equation that explicitly takes into account the role played by
country-specific price indices, called multilateral resistance terms (MRT). They are usually proxied by fixed effects (Feenstra et al., 2005), despite new works propose new solutions:  Baier and Bergstrand (2009) log-linearize the multilateral resistance terms using a first-order Taylor series approximation. This method is termed bonus vetus OLS (BV-OLS). Behrens et al.(2007) suggest a spatial-autoregressive moving-average specification, which results in consistent estimates of the standard gravity equation parameters. Patuelli et al. propose to use a spatial filtering approach.

\item In conclusion, the use of a panel gravity models offer more modeling possibilities as compared to cross-section or time series framework. They contain more variation and less collinearity among the variables and their use results in a greater availability of degrees of freedom and hence increases efficiency in the estimation. (Elhorst, 2003; Baltagi, 2008). However, the panel gravity model has the disadvantages that trade costs cannot be fully recognized by using the distance, which is time-invariant. In this framework, moreover, the effect of the time invariant variables such as contiguity, common language and common currency is generally cut off.
\end{itemize}

\vspace{5mm}

Said that, lets imagine to have a formulation of the spatial gravity model in such a way we account the trade costs using geographical distance and other dummy variables (contiguity, language, currency) as a proxy; we account the spatial effect through a weighted matrix constructed on inverse distance \footnote{Nevertheless, different approaches were proposed in the literature depending on the topic and the purposes of the analysis: the contiguity approach, the technological similarity approach, etc.}, and the multilateral resistance terms through the fixed effects. Two issues follow from this standard procedure: the first is related to the collinearity emerging if we use distance twice, that produce larger standard error of the estimated coefficients;  the second regards the biasness of the distance if used as a proxy of the trade costs in a panel framework. This biasness emerges for several reasons: it is related to 1) the changing nature of the trade costs that distance cannot account for\footnote{trade costs change over time}, 2) the unsuitable choice of the centroid used to define distance, 3) the fact that different kind of transport can have different costs. This last fact is widely argumented in the literature (Head, Mayer 2002, Martinez-Zarzoso, Suarez-Burguet 2005, Anderson, Van Wincoop 2004, Cheng, Wall 2005). Moreover, using fixed effects, it's likely to eliminate the effect of time invariant variables such as contiguity, language, currency. 

The above motivations justify us to test for the possibility to remove the distance variable from the spatial gravity model of trade (to avoid that an improper use of the fixed effects lets the results of the analysis be invalidated), which is the purpose of this work.

Cheng and Wall adopted for this scope a nested approach using Likelihood ratio (LR) tests and they find that fixed effects are suitable to use instead of distance. This approach is open to criticism, since it is not theoretically consistent with the economic derivation of the gravity of trade, and because it doesn't take into account the dispute on how to estimate multilateral resistance terms \footnote{the MRT's are also  traditionaly proxyed by fixed effects, see Feenstra et al.(2005)}. 

A theoretically consistent procedure (which is despite devoted to a little bit different goal) is indeed proposed by Anderson, Yotov (2012).
We are going to check on the possibility to jointly proxy the \emph{multilateral resistance terms} and the \emph{transport costs} (standardly proxied by distance) with a $set$ of fixed effects. Doing in this fashion, we also fit in the dispute about how to estimate the multilateral resistance terms.  

On the light of the Anderson, Yotov (2012) procedure, our proposal is about determining a \emph{multilateral resistance terms} estimated coefficients, through the derivation of the structural model derived by Anderson, Van Wincoop (2003), jointly with an estimation of the empirical fixed effects from an empirical model which doesn't include distance as a proxy for the transport costs. In other words, these coefficients will be determined so that the fixed effects account for the MRT's and the distance effects jointly. Therefore, they represent a suitable proxy for MRT's and distance jointly.  Under such hypotesis, distance and multilateral resistance terms should be equal to the fixed effects unless for a random effect distributed as a white noise (i.e. normal with zero mean).

We will perform on a 22-year OECD panel an ANOVA analysis to study for the MRT's plus distance variance explained by fixed effects. Moreover, a bootstrap analysis to test for the normal distribution with zero mean of the residuals calculated as the difference between MRT's plus distance and the fixed effects.

The results of the just mentioned tests confirm our hypotesis that we can remove the distance if we use a suitable set of fixed effects.

In the next paragraph we discuss the motivation underlying this work. It follows a theoretical section in which we show the framework and how to derive the estimated structural components object of comparison for our analysis. Then, in the fourth section, we show some ad-hoc suitable test and in the fifth section we present an empirical analysis based on a panel of OECD countries in order to check for the above discussed theories. The last section conclusions will be discussed.

\section{Motivations}

We discussed, in the introduction, about the standard procedure used in the spatial gravity model of trade: this model usually account for MRT's with fixed effects, considers the distance as a proxy of the trasport costs (that are time - varying), and it considers the distance twice, if we use the inverse distance approach for the weight matrix.   
Some issues follow from this standard procedure: the first regards the biasness of the distance if used as a proxy of the transport costs, the second is related to the collinearity emerging if we use distance twice.

Starting from the first issue, we said about the adoption in the spatial model of the contiguity matrix $W$ constructed through the inverse of the distance. We are going to have the distance variable used twice in the model. Despite this is, as we already said, a standard practice in the literature, evidences of collinearity aspects are likely to be among the explanatory variables. Collinearity among explanatories determines an increase of the standard errors of the estimated coefficients.

Relating to another issue, it is widely argumented in the literature (Head, Mayer 2002, Martinez-Zarzoso, Suarez-Burguet 2005, Anderson, Van Wincoop 2004) that distance variable is not a suitable proxy in accounting for the \emph{bilateral resistance} effects that the theoretical gravity model identify with trade costs (Anderson, Van Wincoop, 2003). Martinez-Zarzoso and Suarez - Burguet (2005) highlighted that trade costs would be inversely correlated with commercial flow and should be modeled together in a two equations system. This changeable nature of the trade costs make the distance variable inappropriate to proxy the \emph{bilateral resistance}, since it doesn't change over time.

Moreover, using distance we suffer of missmeasurement, since transport costs could change along different type of transport. For example the cost to overland goods by sea is much cheaper than overland goods by land (Cheng, Wall, 2005). Distance is not able to account for this fact.

Another aspect stressed by Cheng and Wall is related to the choice of the centroids in measure the distance between the two countries: sometimes, the choice of the point in space in which we choose the centroid do not fit with the economic center of the countries (in which the main part of the commercial flows happens).
Furthermore, it is empirically demonstrated that the introduction of the distance variable in the model overestimates the negative effect of the \emph{bilateral resistance} (Rose 2002, Cheng, Wall 2005).

One solution could be to use the transport costs to account for itself: in this case the problem disappear. Nevertheless, data on transport costs are hard to obtain: this is the reason why usually we account this by distance proxy.

The above motivations justify us to test for the possibility to remove the distance variable from the gravity (to avoid that an improper use of the fixed effects lets the results of the analysis be invalidated), which is the goal of this work.

Cheng and Wall adopted for this scope a nested approach using Likelihood ratio (LR) tests. They compare a model with distance and a model with fixed effects. This approach is open to criticism, since it is not theoretically consistent with the economic derivation of the gravity of trade, and because it doesn't take into account the dispute on how to estimate multilateral resistance terms. Infact, the MRT are also usually estimated by fixed effects. 

A theoretically consistent procedure is indeed proposed by Anderson, Yotov (2012), which is based on the economic derivation of the structural gravity model by Anderson, Van Wincoop (2003).
Fitting in the dispute about how to estimate the multilateral resistance terms \footnote{Feenstra et al. (2005), Baier Bergstand (2009), Behrens, Ertur, Koch (2007), Patuelli et al.}, we are interested to check on the possibility to jointly proxy the \emph{multilateral resistance terms} and the \emph{transport costs} (standardly proxied by distance) with a $set$ of fixed effects.  

On the light of the Anderson, Yotov (2012) procedure, our proposal is about determining a \emph{multilateral resistance terms} estimated coefficients, through the derivation of the structural model derived by Anderson, Van Wincoop (2003), jointly with an estimation of the empirical fixed effects from an empirical model which doesn't include distance as a proxy for the transport costs. In other words, these coefficients will be determined so that the fixed effects account for the MRT's and the distance effects jointly. Therefore, we wish they represent a suitable proxy for MRT's and distance jointly.

\section{Theoretical procedure to derive the estimated structural terms}

To derive the estimated coefficients we spoke about before, we make use of the derivation of the structural model by Anderson, Van Wincoop (2003).

To derive this model we adopt the concept of trade separability: for this concept, at an upper level we generate the production  and the expenditure for each good in each region, and at a lower level the demand and supply, conditioned to the production and expenditure value determined before.

To derive the model we make use of the following constraints:
\begin{itemize}
  \item $\emph{budget constraint}$ ( one for each destination);
  \item $\emph{market clearance condition}$ (one for each origin);
\end{itemize}

jointly with the use of the $CES$ demand function.

Defining $T_{ij} \geq 1$ as the variable relative to the whole trade costs and $\sigma$ the elasticity of substitution, we define the $CES$ demand function.

\begin{equation}\label{eulero}
Y_{ij} = (\beta_i p_i T_{ij} / P_j)^{1-\sigma} E_j
\end{equation}

The budget constraint determines on the relative price:
\begin{equation}\label{eulero}
P_{j} = [\sum_i \beta_i p_i T_{ij})^{1-\sigma}]^{1/1-\sigma}
\end{equation}

Assuming the $\emph{market clearance condition}$ , in other words $X_i = \sum_j X_{ij}$, we sum over $j$ the $CES$ demand function to obtain that:

\begin{equation}\label{eulero}
  X_i = \sum_j (\beta_i p_i)^{1-\sigma} (T_{ij}/P_j)^{1-\sigma} E_j.
\end{equation}
Using $X = \sum_i X_i$, we obtain that
\begin{equation}\label{eulero}
  X_i / X = (\beta_i p_i \Pi_i)^{1-\sigma}
\end{equation}
where $\Pi_i = \sum_j(T_{ij} P_j)^{1-\sigma}E_j / X$

For the $\emph{market clearence condition}$ constraint, if we sum over $i$ the last equation we obtain that $\sum_i(\beta_i p_i \Pi_i)^{1-\sigma} = 1 $.

Using the same equation to substitute his terms $p_i$ e $\beta_i$ in the initial demand function , to obtain a three equation system representing the gravity model:

\begin{equation}\label{eulero}
  Y_{ij} = \frac{E_j X_i}{X} (\frac{T_{ij}}{P_{j} \Pi_{i}})^{1-\sigma}
\end{equation}
\begin{equation}\label{eulero}
 (\Pi_{i})^{1-\sigma}= \sum_{j}(\frac{T_{ij}}{P_{j}})^{1-\sigma}(\frac{E_{j}}{X})
\end{equation}
\begin{equation}\label{eulero}
   (P_{j})^{1-\sigma} = \sum_{i}(\frac{T_{ij}}{\Pi_{i}})^{1-\sigma}(\frac{X_{i}}{X})
\end{equation}

Hence, the gravity model derived by Anderson and Van Wincoop considers the \emph{multilateral resistance effects}, which are unknown, but we can estimate them through a three equation system defined above.

The equations $(5)-(7)$ are fondant for our analysis: their stochastic versions permit us to estimate the theoretical $\emph{multilateral resistance effects}$. These estimates are based on a previous estimate of the $E_j$, $X_i$ and $T_{ij}$ components, where each of these will be defined on the basis of empirical values, as we will see.

$$ $$
We can estimate the empirical values trough the equation (8).

\begin{equation}\label{eulero}
Export_{ij} = (\hat{X}_{i}) + (\hat{E}_{j}) + (\hat{T2}_{ij}) + (others_{ij}) + \theta_{ij} + \varepsilon_{ij}
\end{equation}
$$ $$
where  $\theta_{ij}$ should account for the effect of the transport costs and of the \emph{multilateral resistance terms}. Moreover, with $\hat{X}_{i}$ we refer to an estimated value for the reporter economic dimension, with $\hat{E}_{j}$ we refer to an estimated value for the partner economic dimension, and $\hat{T2}_{ij}$ represent the estimated trade costs different from transport costs (total trade costs $-$ transport costs)
$$ $$
This step, in fact, is twofold:  we estimate with the same equation model: a) a set of  $n*n$ fixed effects defined such that they account for the effect of the \emph{multilateral resistance} and the effect of transport costs, b) the estimated structural terms of the gravity.

The choice of the right set of fixed effects is quite relevant: the distance accounts for the transport costs between couple of country, which could be different along different directions (from $i$ to $j$ or from $j$ to $i$). So, to use a $set$ of symmetric effects could not be suitable. At the same, to consider two $sets$ of fixed effects, one for the $reporter$ country, the other for the $partner$ country, could be worse, since couple fixed effects account for much more variability.

One aspect of debate regards the choice to identify the distance as the only proxy variables of trade costs we want to remove. That is because we assume the other proxies for trade costs:

\begin{itemize}
  \item unbiased and without identification problem;
  \item of standard use in the model.
\end{itemize}

Hence, assuming this hypotesis, we need for an evidence of the decomposability of trade costs. Feenstra (1998) shows that the total trade costs are due for the $21\div$ to the transport costs (for which distance is used as a proxy). The other part of the cost are motivated by the market entry barriers.

We can assume $\hat{T}_{ij} = dist_{ij} + \hat{T2}_{ij}  $ be the equation defining the total trade costs decomposition in transport costs and other \emph{bilateral resistance} costs ($T2$).

To resume, we determine by a stochastic version of the equations $(5)-(7)$, with support of (8), the estimated coefficients for \emph{multilateral resistance terms} $[\widehat{\Pi}_i; \widehat{P}_j]$, and the estimated set of couple fixed effect.

\section{Proposed analysis and tests}

The structural model theory would implies that the expected value of the estimated fixed effect would be equal to the expected value of the estimated \emph{multilateral resistance terms} plus distance. Differences might only be due to random deviation around the zero (i.e. the difference is distributed as a white noise):

 \begin{equation}\label{eulero}
   \theta_{ij} = \Pi_i +  P_j  +  dist_{ij}. \footnote{Here, the fact that we do not have hat indicates that we think this relation at an expected value level}
 \end{equation}

It is possible to test the above mentioned equation by comparing those estimated components:
\begin{equation}\label{eulero}
 r_{ij}= \widehat{\theta}_{ij} - (\widehat{\Pi}_i + \widehat{P}_j + dist_{ij}).
\end{equation}

The $r_{ij}$ term could be interpreted as a residual component of a regression model of the estimated fixed effects $\widehat{\theta}_{ij}$ respect to $\widehat{\Pi}_i + \widehat{P}_j +  dist_{ij}$, with the constraint that the intercept must be equal to $0$ and every slope parameter is equal to $1$ .

That interpretation permit us to analyze the $r_{ij}$ component through an analysis of variance ($ANOVA$, as made by Egger, Pfaffermayr (2002, 2004), as a fitting measure for the structural components. The ANOVA is considered as a special case of the linear regression model (Gelman 2005, Montgomery 2001). Both, indeed, consider the observed $\theta_{ij}$ as a summation of the structural variables and the residuals.

The statistic significance of the $test$ is determined by the ratio between the two variance. In other words, placing:
  \begin{itemize}
    \item $V(r)$  the variance of $r_{ij}$,
    \item $V(S)$ the variance of $\widehat{\Pi}_i + \widehat{P}_j + dist_{ij}$;
  \end{itemize}

the proportion of the variance of the fixed effects $\widehat{\theta}_{ij}$ explained by the components of the structural model $\widehat{\pi}_i  +  \widehat{P}_j + \widehat{dist}_{ij}$ will be $\frac{1-V(r)}{V(S)}$, which is the correlation index $R^2$ of the regression model if we assume the above constraints.

To clarify, with this test we want to evaluate how much variability of the fixed effects is explained by their theoretical  counterpart (i.e. the MRT's used on the structural model and the distance as a proxy of transport costs).

 Moreover, a further hypotesis must be assumed: the hypotesis of indipendence  of test from possible alteration of the experiment.  This hypotesis is satisfied, because both the dependent and the explanatories  in the $ANOVA$ test were defined conditionally to the same structural variables composing the model.

\vspace{5mm}

The interest for the validation of the empirical model without distance need for a further suitable test for the null hypotesis regarding the expected value of  the $r_{ij}$'s.

Unfortunately, an hypotesis test like the $t-test$ to evaluate the null $H_0: E(r_{ij})=0$ needs the assumption that such residuals are generated from an indipendent and identically distributed (i.i.d.) process, and that the sample mean of the $r_{ij}$'s is normally distributed.  The second assumption in ensured from the result of the central limit theory.

Instead, the i.i.d. assumption is hardly conceivable, in fact, such residuals are generated from a system that jointly considers the \emph{multilateral resistance terms} ($\Pi_i$ e $P_j$) and the distance ($dist_{ij}$), which components are each other correlated: in fact, the \emph{multilateral resistance terms} are estimated through the equations (6) and (7) using the distance as true value of the transport cost among the regressor, which contains a measurement error (since it is a proxy). Such measurement error affects the estimated \emph{multilateral resistance terms}, that will be correlated with the distance.

A possible solution could be to perform a $t-test$ based on $bootstrap$ procedure, bootstrapping directly the $r_{ij}$ residuals. This procedure presents some problems, in fact, the bootstrap procedure need for the i.i.d. assumption of the bootstrapped residuals. Since we are going to have dependent residuals, a solution proposed in the literature could be to perform the $block-bootstrap$. However, this approach need to vary the dimension of the sub-samples (see Politis, Romano, 1994).

Conversely, is possible to assume that the $\varepsilon_{ij}$ residuals from model in (8) are i.i.d.. Through a \emph{regression bootstrap} procedure which use the residuals in (8) as a starting point, is possible to obtain a set of $r_{ij}$ $bootstrapped$ residuals  which are i.i.d..

Here, we describe the step by step procedure.

 \begin{enumerate}
   \item To estimate (8) and to obtain the estimated $\varepsilon_{ij}$;
   \item To perform a $bootstrap$ on the residuals generated at the first step:  we are going to obtain $[\varepsilon_{ij,1}, ..., \varepsilon_{ij,B}]$. We compute a number equal to $B$ $bootstraped$ terms for $Export_{ij}$:

       ([$Export_{ij,1}, ..., Export_{ij,B}$]) through the following relation:

       $Export_{ij,b} =  \widehat{Export}_{ij} + \varepsilon_{ij,b}; \forall b = 1,...,B$

       where $\widehat{Export}_{ij}$ is the estimation from the model (8);
   \item To perform a number of $B$ linear regressions of the model (8), using, for each one, each of the $B$ $bootstraped$ terms for $Export_{ij}$ obtained at the second step; in order to obtain $B$ sets of coefficients for the model (8) and $B$ estimations for the fixed effects $\theta_{ij}$;
   \item For each of the $B$ iterations, to re-estimate the \emph{multilateral resistance terms} throughout the relations defined in (6) and (7).
 \end{enumerate}

At the end of this procedure, we would have $B$ $bootstraped$ estimates for the fixed effects: $[\theta_{ij,1}, ...,\theta_{ij,1}]$. $B$ $bootstraped$ estimates for the \emph{multilateral resistance terms}: $[\Pi_{i,1}, ..., \Pi_{i,B}]$, $[P_{j,1}, ..., P_{j,B}]$. $B$ estimates for the distance: $[dist_{ij,1}, ..., dist_{ij,B}]$.  To the end, we can define $B$ $bootstraped$ i.i.d. terms representing the $r_{ij}$'s, so defined:
$$ $$
 $r_{ij,b} = \theta_{ij,b} - ( \Pi_{i,b} + P_{j,b} + dist_{ij,b}); \forall b = 1, ..., B$
$$ $$
Now, to compute a $t-test$ become possible. For each iteration we compute the mean:

\begin{equation*}
 M(r_{ij,b}) = \frac{\sum_{ij} r_{ij,b}}{n^2},
\end{equation*}
And its $\emph{standard error}$:

\begin{equation*}
 SE(r_{ij,b}) = [\frac{\sum_{b=1}^{B} M(r_{ij,b} - M(r_{ij})}{\sqrt{n^2}}]^{-1/2},
\end{equation*}
 where $M(r_{ij}) = \sum_{i=1}^{B}M(r_{ij,b})$ represents the mean of the means.

The $t-test$  to evaluate the $H_0: r_{ij} = 0$ hypotesis  assumes the following form:

\begin{equation}\label{eulero}
t \sim \frac{M(r_{ij})}{SE(r_{ij,b})}
\end{equation}

\section{Empirical Analysis}

\subsection{Choice of the empirical model}

To test the proposed analyzes we  use a panel related to the country-to-country trade flow over the period from 1988
to 2009 for the OECD members. The sample contains 32 countries, and the time series is 22 years long, resulting in a $n*n*T$ = 22528 observations.

As the explanatory variables of the empirical gravity model we use the  consumer price index at purchase power
of parity (ppp), the gross domestic product (gdp) in real terms and the population
(pop) for the origin and for the destination country. We use distance among centroids to proxy for the transport costs.  Moreover, besides the dummies contig, comcur, and comlang, we use the
dummies relatives to the free trade agreement of EU 15, NAFTA and EFTA. Furthermore, the dummy variables representing the
 effect of each couple of countries are inserted in the
model and the stock of immigrants. The purchase power parity index is available at PennTable web site, in
a dataset called PWT 7.0. The explanatory variable, as well, are available at
the database of PennTable (gdp and pop), and from CEPII (comlang, contig
and comcur).

Along with the spatial effects components, we define a weight matrix using the inverse of the distance between the centroids (available at CEPII as well).

As proposed by Arbia (2009) we check the spatial effects assumption by mean of the
Moran I on the OLS residuals with the inverse distance matrix. This test show
a significant spatial effect. We choose the SAR model in which the spatial lag of the dependent variable is taken into account \footnote{We motivate this choice because we are interested in taking into account the spatial spillover effect. Le Sage, Pace (2008) developed a SAR model consistent with this effect.}. Nevertheless, also the Durbin  \footnote{Derived to account for spatially correlated missing variables and/or common shocks (Le Sage,Pace, 2008).}  is estimated, in order to make the results consistent. A spatial Hausmann test was performed, in order to choice between fixed and random effects. This test highlights the strong preference for the fixed effects model. Furthermore, a Spatial Lagrange Multiplier test was performed to check for the presence of spatial autocorrelation. 

Since we have a spatial autoregressive model, we need to control for the intrinsic endogeneity of the explanatories. We do so by using IV/GMM techniques (Kelejian, Prucha 1998, 1999; Mutl, Pfaffermayr, 2007).

Therefore, based both on our tests and on the literature, we have reasons to choose a spatial model where distance appears twice (as proxy for transport costs and to define the weighted matrix):  this follows the benchmark gravity models of trade, where the spatial effects is taken into account through a weighted matrix constructed on inverse distance.

The empirical model takes therefore the following form:

  \begin{align}
 Export_{ijt}  &=  \alpha_{ij} + \alpha_t + \beta^o_{1}Pop^o_{it} + \beta^o_{2} GDP^o_{it} + \beta^o_{3} ppp^o_{it} + \beta^o_{4}Nafta^o_{it} + \beta^o_{5}  Efta^o_{it} + \nonumber \\ 
& \beta^o_{6} Eu15^o_{it} + \beta^d_{1}Pop^d_{jt} + \beta^d_{2} GDP^d_{jt} + \beta^d_{3} ppp^d_{jt} + \beta^d_{4}Nafta^d_{jt} + \beta^d_{5}  Efta^d_{jt}  + \nonumber \\ 
& + \beta^d_{6} +  Eu15^d_{jt}\beta_7 Migrat_{ijt} + \psi^{od}_{1} Contig_{ij} + \psi^{od}_{2} Comlang_{ij} + \psi^{od}_{3} Comcur_{ij}  + \nonumber \\  
& +  \psi^{od}_{4} Dist_{ij} \psi_5 w_{ij,hk,OD} bilat.PIL_{ijt} + \rho w_{ij,hk,OD} y_{hk,t}  + u_{ijt}; \nonumber \\  
  \end{align}

\subsection{analysis and results}

We can apply the analyzes proposed in in the previous section at theoretical level. First of all, we need to define the empirical model without distance specified in (8): we only  removed the variable $dist$ from (12):

  \begin{align}
 Export_{ijt}  &= \alpha_{ij} + \alpha_t + \beta^o_{1}Pop^o_{it} + \beta^o_{2} GDP^o_{it} + \beta^o_{3} ppp^o_{it} + \beta^o_{4}Nafta^o_{it} + \beta^o_{5}  Efta^o_{it} + \nonumber \\ 
&  \beta^o_{6} Eu15^o_{it} + \beta^d_{1}Pop^d_{jt} + \beta^d_{2} GDP^d_{jt} + \beta^d_{3} ppp^d_{jt} + \beta^d_{4}Nafta^d_{jt} + \beta^d_{5}  Efta^d_{jt}  +  \nonumber \\ 
&+ \beta^d_{6} Eu15^d_{jt} + \beta_7 Migrat_{ijt} + \psi^{od}_{1} Contig_{ij} + \psi^{od}_{2} Comlang_{ij} + \psi^{od}_{3} Comcur_{ij} +  ; \nonumber\\
& \psi_5 w_{ij,hk,OD} bilat.PIL_{ijt} + \rho w_{ij,hk,OD} y_{hk,t}  + u_{ijt} \nonumber \\
  \end{align}

In this equation, we've got the variable we must use to replace the terms of economic dimensions for partner and reporter countries, in order to obtain the estimated $MRT'$s via stochastic version of (6) and (7):

  \begin{itemize}
\item $\hat{X}_{it}=  \beta^o_{1}Pop^o_{it} + \beta^o_{2} GDP^o_{it} + \beta^o_{3} ppp^o_{it}$
\item $\hat{E}_{jt} = \beta^d_{1}Pop^d_{jt} + \beta^d_{2} GDP^d_{jt} + \beta^d_{3} ppp^d_{jt}$
  \end{itemize}

Moreover, the estimation for the trade costs $T_{ij}$ comes from this proxy:

\begin{itemize}
\item $\hat{T}_{ij} = \psi^{od}_{1} Contig_{ij} + \psi^{od}_{2} Comlang_{ij} + \psi^{od}_{3} Comcur_{ij} + \psi^{od}_{4} Dist_{ij}$
\end{itemize}

\vspace{5mm}

We have to note that, with this application, we manage with panel data and with the $t$ index for the time series. This permits us to obtain the estimation of the $n*n$ fixed effects, which is not possible for cross section, since the sample dimension is $n*n$. However, engaging a panel data analysis, we need to introduce some implementation to the theory showed in the previous section, in order to identify the estimated $MRT'$s.
It is straightforward to implement the proposed methods in the previous sections for the case of panel data by adding summation over $t$ in equations (6) and (7).

\vspace{5mm}

Proceed to an ANOVA analysis as described in the theory, with the aim to investigate how much of the fixed effects is explained by the structural variables.

The result is a $R^2$ of $0.9998$\footnote{the ANOVA results based on DURBIN model are available upon request}.

This value, surprisingly high, tells us that the $99.98 \div$'s of the total variance of the $\theta$'s is explained by the structural model theorized by Anderson and Van Wincoop (2003), and gives us some demonstration of the fact that:
\begin{enumerate}
\item the structural model theorized by Anderson and Van Wincoop is a consistent formulation for the international trade phenomenon;
\item the effect of the distance can be accounted by a set of couple fixed effects;
\end{enumerate}

\vspace{5mm}

Proceed to a Bootstrap analysis as described in the theory, now with the purpose to test if the expected value of $r_{ij}$ is equal to zero.

The result of the $t-test$ showed in equation (11) assumes value equal to $1.12$, which have a corresponding $p-value$ of  $Pr(| T |> | t |)=$ 0.130, therefore, the null hypotesis is not \footnote{the bootstrap results based on DURBIN model are available upon request}.
This analysis confirms the thesis for which the empirical gravity model can be consistently estimated without the inclusion of the distance variable, if properly replaced by couple fixed effects.

\subsection{comparison and remarks}

At an explorative level, to investigate the consistency of the previous analyzes, we are interested to compare the estimated coefficients of the two $SAR$ models \footnote{the comparison between the DURBIN models are available upon request}, the first one with distance, the second without distance. Remember that the aim was to remove the distance from the model and the tests confirmed that this is possible. What happen to the estimated coefficients, once we remove distance variable from the model? We note that several variables increase their coefficients once the distance variable do not take part in the model (as we can see in table 1): the effects of the distance is therefore caught by the other variables. Nevertheless, the signs of the coefficients of the model without distance are in line with the literature.

\begin{table}
\caption{A preliminary comparison:  $IV/GMM$ estimated coefficients for SAR models with and without distance}
\center
\begin{tabular}{|c|c|c|}
\hline			
VARIABLE  &  without distance (*) &  with distance  \\
\hline
$\alpha$ &  - & - \\		
Contig &	1.297 ($0.033$)*** &	0.395 ($0.029$)  \\
Comlang  &	0.877 ($0.033$)*** &	1.084 ($0.028$)  \\
Dist &	-  &	-0.682 ($0.078$)*** \\
Comcur &	3.335 ($0.151$)***& 0.549 ($0.128$)***  \\
Popd  &	1.509 ($0.037$)&	0.937 ($0.029$)***  \\
pppd  & 0.289	($0.014$)*** & 0.095 ($0.123$)***  \\
GDPd & 2.679 ($0.078$)*** & 1.212 ($0.064$)***  \\
Popo  & 1.793	($0.051$)***	& 0.953 ($0.043$)***  \\
pppo & 0.445	($0.019$)*** & 0.159 ($0.016$)***  \\
GDPo  & 2.484	($0.076$)***& 1.170 ($0.065$)***  \\
Eftao  & 2.034 ($0.084$)*** & 1.9269 ($0.071$)*** \\
Eftad  & 0.312 ($0.035$)*** & 0.3981 ($0.031$)***\\
Naftao  & -2.743 ($0.134$)*** & -2.533 ($0.143$)***\\
Naftad & -2.117 ($0.089$)*** & -1.993 ($0.076$)*** \\
EU15o  & 0.830 ($0.026$)*** & 0.7937 ($0.022$)*** \\
EU15d  & 0.773 ($0.026$)*** & 0.7236 ($0.018$)*** \\
W Bilat. PIL & -0.108 ($0.379$)*** & -0.080($0.347$)*** \\
Migrat &  0.019 ($0.051$)*** & 0.032 ($0.043$)***\\

Rho	& 0.00055 *** &	0.00035 *** \\
\hline
\end{tabular}

\vspace{5mm}
\begin{flushleft}
* In brackets the standard error
\end{flushleft}
\end{table}

\section{Conclusions and future developments}

As previously motivated, the use of the $distance$  variable have a critical role when we work with spatial gravity of trade in a panel framework.  Recent literature  address this issue replacing with the fixed effects the distance used as a proxy for the transport costs. It was motivated the reason why the couple fixed effects without the hypothesis of symmetry are preferred to the couple fixed effects with hypothesis of symmetry; it was also motivated why fixed effects for reporter and partner countries are not suitable to proxy the transport costs. The analysis carried out here is based on the theoretical definition of the unconstrained gravity model, and  it is consistent from an economic point of view. The analysis is based on a comparison between the estimated fixed fixed and the distance (plus the multilateral resistance terms estimated from the theoretical model).

\vspace{4mm}

The analyzes show interesting results: the first is that the structural model derived by Anderson, Van Wincoop seems to be consistent, since the theoretical $MRT$'s explain almost the total fixed effects variance. The second regards the confirmed possibility to remove the distance from the model: the effect of the $MRT$'s and of the transport cost is excellently grasped by a set of couple non-symmetric fixed effects.

\vspace{4mm}

The panel data application is very important: it allows us to use and estimate a set of $n^2$ fixed effects. This kind of fixed effects permit us to consider (and to proxy) more carefully the impact of transport costs, because we have one cost for each couple of countries, and they are not symmetric (costs from $i$ to $j$ $\neq$ costs from $j$ to $i$).

\vspace{4mm}

A further development could be to investigate if at a sub-sectorial level this results still hold.

\section{Bibliography}

\begin{itemize}
\item Adam C., Cobham D. (2007): Modelling multilateral trade resistance in a gravity model with exchange rate regimes. Centre for dynamic macroeconomic analysis conference papers 2007.
\item Anderson J.A. (1979): A Theoretical Foundation for the Gravity Model, The American Economic Review, 69,106-116.
\item Anderson J.A., Van Wincoop E. (2003): Gravity with Gravitas: A solution to the Border Puzzle, American Economic Review 93, 170-192.
\item Anderson J.A., Van Wincoop E. (2004): Trade Costs. Journal of economic literature, 42, 691-751.
\item Anderson J.A., Yotov Y.V. (2008): The changing incidence of geography. American Economic Review, vol. 100(5), p.2157-86.
\item Anderson J.A., Yotov Y.V. (2010): Specialization: Pro and Anti-globalization, 1990-2002. NBER Working Paper No. 16301.
\item Anderson J.A., Yotov (2012): Gold Standard Gravity, NBER Working Papers 17835, National Bureau of Economic Research, Inc.
\item Anselin L. (1988): Spatial econometrics: Methods and Models (Kluwer Academic Publishers, Dordrecht)
\item Baier S.L. and J.H. Bergstrand (2009) Bonus Vetus OLS: A Simple Method for Approximating International Trade-Cost Effects Using the Gravity Equation. Journal of International Economics 77 (1), 77-85
\item Baldwin R., Taglioni D. (2006): Gravity for Dummies and Dummies for gravity equation - NBER Working Paper No. 12516.
\item Baltagi BH (2008): Econometric analysis of Panel Data, 4th edition, John Wiley e sons. Ltd.
\item B. H. Baltagi, P. Egger, and M. Pfaffermayr, ?Estimating models of complex fdi: Are there third-country effects?? Journal of Econometrics, vol. 140, no. 1, pp. 260-281, 2007.
\item J. T. Bang, Regional integration, spatial effects, and the gravity equation, 2006.
\item Behrens K., C. Ertur and W. Koch (2007) 'Dual' gravity: Using spatial econometrics to control for multilateral resistance (CORE Discussion Papers No. 2007059). Louvain: Universit� catholique de Louvain, Center for Operations Research and Econometrics (CORE)
\item Egger P., Pfaffermayr M. (2002-1): The prime effect of the european integration intra EU core and periphery trade. Working paper in econometrics, University of Innsbruck.
\item Egger P., Pfaffermayr M. (2002-2): Foreign direct investment and european interaction in the 90's. Working paper in econometrics, University of Innsbruck.
\item Egger P., Pfaffermayr M. (2004): The impact of bilateral investment treaties on foreign direct investment, Journal of Comparative Economics, Volume 32, Issue 4, Pages 788-804, ISSN 0147-5967, 10.1016/j.jce.2004.07.001.
\item Elhorst, J. P. (2003). Specification and estimation of spatial panel data models. International regional science review, 26(3), 244-268.
\item Everett, S.J., Hutchinson, W.K. (2002): The Gravity equation in international economics. Scottish Journal of political economy, vol.49, no.5.
\item Feenstra C.R. (1998): Integration of Trade and Disintegration of Production in the Global Economy. The Journal of Economic Perspectives, Vol. 12, No. 4, pp. 31-50.
\item Feenstra R.C., R.E. Lipsey, H. Deng, A.C. Ma and H. Mo (2005) World Trade Flows: 1962- 2000 (NBER Working Paper). Cambridge: National Bureau of Economic Research
\item Fischer, M. M., Griffith, D. A. (2008). MODELING SPATIAL AUTOCORRELATION IN SPATIAL INTERACTION DATA: AN APPLICATION TO PATENT CITATION DATA IN THE EUROPEAN UNION*. Journal of Regional Science, 48(5), 969-989.
\item Gelman A. (2005): Analysis of variance - why it is more important than ever. Ann. Statist. Volume 33, Number 1, 1-53.
\item Gorman, S. P., Patuelli, R., Reggiani, A., Nijkamp, P., Kulkarni, R.,  Haag, G. (2007). An application of complex network theory to German commuting patterns. In Network Science, Nonlinear Science and Infrastructure Systems (pp. 167-185). Springer US.
\item S. G. Hall and P. Petroulas, Spatial interdependencies of FDI locations: A lessening of the tyranny of distance? Bank of Greece, 2008, vol. 67.
\item Head K., Mayer T. (2002): Illusory Border Effects Distance missmeasurement inflates estimates of home bias in trade.
\item Kang, Heejoon and Fratianni, Michele U. (2006): International Trade Efficiency, the Gravity Equation, and the Stochastic Frontier-
\item Kapoor M., Kelejian H.H., Prucha I.R. (2007): Panel Data Models with Spatially Correlated Error Components. Centre for analytical Finance Research Paper No.02-05.
\item Kelejian H.H., Prucha I.R. (1998): a Generalized Spatial two-stage least squares procedure for estimating a spatial autoregressive model with autoregressive disturbances. Journal of real estate finance and economics, Vol.17:1,99-121.
\item Kelejian, H. H., Prucha, I. R. (1999). A generalized moments estimator for the autoregressive parameter in a spatial model. International economic review, 40(2), 509-533.
\item H. Kelejian, G. S. Tavlas, and P. Petroulas, In the neighborhood: The trade effects of the euro in a spatial framework, Regional Science and Urban Economics, vol. 42, no. 1, pp. 314-322, 2012.
\item LeSage J. and Pace R.K., (2008): Spatial econometrics modeling of origin-destination flows. Journal of Regional Science, 48(5): 941967
\item Martinez Zarzoso I., Suarez Burguet C., (2005): Transport costs and trade: Empirical evidence for latino American imports from the European union. The Journal of International Trade and Economic Development: Vol. 14, Iss. 3.
\item Mutl, J., Pfaffermayr, M. (2008). The spatial random effects and the spatial fixed effects model: the Hausman test in a Cliff and Ord panel model (No. 229). Reihe �konomie/Economics Series, Institut f�r H�here Studien (IHS).
\item Mutl J., Pfaffermayr M. (2008): The Hausman Test in a Cliff and Ord Panel Model. Econometrics Journal, volume 10, pp.1-30.
\item Patuelli R., Linders GJ, Metulini R., Griffith DA: The Space of Gravity: Spatial Filtering Estimation of a
Gravity Model for Bilateral Trade (Working paper).
\item Politis, Romano (1994): The Stationary Bootstrap. Journal of the American Statistical Association, Volume 89, Issue 428.
\item Rose, A.K., (2000): One money, one market: Estimating the effect of common currencies on trade. Economic Policy 20 (April), 7 - 45.
\item Tinbergen J. (1962): Shaping the World Economy; Suggestions for an International Economic Policy - Twenty Century Fund, New York.
\item Tobler W.R. (1970): A computer movie simulating urban growth in the Detroit region. Economic Geography, vol. 46.

\end{itemize}

\end{document}